\documentclass[aps,pre,twocolumn,groupedaddress,]{revtex4}
\usepackage{graphics,color}

\usepackage[]{graphicx}
\DeclareGraphicsExtensions{.pdf}

\begin{document}

\title{Thermodynamics for a network of neurons: Signatures of criticality}

\author{Ga\v{s}per Tka\v{c}ik,$^a$ Thierry Mora,$^b$ Olivier Marre,$^c$ Dario Amodei,$^{d,e}$ Michael J. Berry II,$^{e,f}$ and William Bialek$^{d,g,h}$}

\affiliation{$^a$Institute of Science and Technology Austria, Am Campus 1, A--3400 Klosterneuburg, Austria,\\
$^b$Laboratoire de Physique Statistique, CNRS, UPMC and  l'\'Ecole Normale Sup\'erieure,  24 rue Lhomond, 75231 Paris Cedex 05, France,\\
$^c$Institut de la Vision, UMRS 968 UPMC, INSERM, CNRS U7210, CHNO Quinze--Vingts, F--75012 Paris, France,\\
$^d$Joseph Henry Laboratories of Physics, 
$^e$Princeton Neuroscience Institute,  
$^f$Department of Molecular Biology, and
$^g$Lewis--Sigler Institute for Integrative Genomics,
Princeton University, Princeton, New Jersey 08544 USA,\\
$^h$Initiative for the Theoretical Sciences, The Graduate Center, City University of New York, 365 Fifth Ave., New York, New York 10016 USA}

\date{\today}

\begin{abstract}
The activity of a neural network is defined by patterns of spiking and silence from the individual neurons.  Because spikes are (relatively) sparse,  patterns of activity with increasing numbers of spikes are less probable, but with more spikes the number of possible patterns increases.  This tradeoff between probability and numerosity is mathematically equivalent to the relationship between entropy and energy in statistical physics.  We construct this relationship for populations of up to $N=160$ neurons in a small patch of the vertebrate retina,  using a combination of direct and model--based analyses of experiments on the response of this network to naturalistic movies.  We see signs of a thermodynamic limit, where the entropy per neuron approaches a smooth function of the energy per neuron as $N$ increases.  The form of this function corresponds to the distribution of activity being poised near an unusual kind of critical point.  Networks with more or less correlation among neurons would not reach this critical state.  We suggest further tests of criticality, and give a brief discussion of its functional significance.
\end{abstract}

\maketitle

\section{Introduction}

Our perception of the world seems a coherent whole, yet it is built out of the activities of  thousands or even millions of neurons, and similarly for our memories, thoughts, and actions.  It seems difficult to understand the emergence of behavioral and phenomenal coherence unless the underlying neural activity  also is coherent.  Put simply, the activity of a brain---or even a small region of a brain devoted to a particular task---cannot  be just the summed activity of many independent neurons.  But if  neurons are not independent, how do we describe their collective activity?

Statistical mechanics provides a language for connecting the interactions among many microscopic degrees of freedom to the macroscopic behavior of matter.    Thus, we use statistical mechanics to give a quantitative theory of how a rigid solid emerges from the interactions between atoms, how a magnet emerges from the interactions between electron spins, and so on \cite{anderson_72,textbook}.    Importantly, these are all collective phenomena:  there is no sense in which a single molecule, or even a small cluster, can be described as  solid or liquid; rather, solid and liquid are statements about the joint behaviors of  many, many molecules.  

At the core of equilibrium statistical mechanics is the Boltzmann distribution, which describes the probability of finding a system in any one of its possible microscopic states.  As we consider systems with larger and larger numbers of degrees of freedom, this probabilistic description  converges onto a deterministic, thermodynamic description.  In the emergence of thermodynamics from statistical mechanics, many microscopic details are lost, and many systems that differ in their microscopic constituents nonetheless exhibit quantitatively similar thermodynamic behavior.  Perhaps the oldest example of this idea is the ``law of corresponding states''   \cite{guggenheim_45}.

The power of statistical mechanics to describe collective, emergent phenomena in the inanimate world led many people to hope that it might also provide a natural language for describing networks of neurons \cite{hopfield_82,amit+al_87,amit_89,hertz+al_91}.   But if one takes the language of statistical mechanics seriously, then as we consider networks with larger and larger numbers of neurons, we should see the emergence of something like thermodynamics.  If we can find this ``thermodynamic limit'' for real neurons, we can hope to find simpler universal behaviors for the network as a whole, independent of many microscopic details.

\section{Theory}
\label{looking_for}

At first sight, the notion of a thermodynamics for neural networks seems hopeless.  Thermodynamics is about temperature and heat, both of which are irrelevant to the dynamics of these complex, non--equilibrium systems.   But all of the thermodynamic variables that we can measure experimentally in an equilibrium system can be calculated from the  Boltzmann distribution, and hence statements about thermodynamics are equivalent to  statements about this underlying probability distribution.   It is then only a small jump to realize that  {\em all} probability distributions over $N$ variables can have an associated thermodynamics in the $N\rightarrow\infty$ limit.   This link between probability and thermodynamics is   well studied by mathematical physicists \cite{ruelle}, and has been a useful guide to the analysis of experiments on dynamical systems \cite{halsey+al_86,feigenbaum+al_86}, although perhaps still is not as widely appreciated as it might be.

To be concrete, we consider a system with $N$ elements, and each element is described by a state $\sigma_{\rm i}$; the  state of the entire system is given by  ${\mathbf \sigma} \equiv \{\sigma_1 ,\, \sigma_2, \,\cdots , \, \sigma_N\}$.  We are interested in the probability $P({\mathbf\sigma})$ that we will find the system in any one of its possible states.   There is one state ${\mathbf\sigma}_0$ that is the most likely state, and we can measure all probabilities relative to the probability of this state.  In particular, we can define
\begin{equation}
E({\mathbf\sigma}) = -\ln \left[ {{P({\mathbf\sigma})}\over{P({\mathbf\sigma}_0)}}\right] .
\label{Edef}
\end{equation}
In an equilibrium system, this is precisely the energy of each state (in units of $k_B T$), but we can define this ``energy'' for any probability distribution.  As discussed in detail in Appendix \ref{thermoP}, all of thermodynamics can be derived from the distribution of these energies.  Specifically, what matters is how many states have $E({\mathbf\sigma})$ close to a particular value $E$.  We can count this number of states, $n(E)$, and define a local (microcanonical) entropy $S(E) = \ln n(E)$.    If we can imagine a family of systems in which the number of degrees of freedom $N$ varies, then a thermodynamic limit will exist provided that both the entropy and the energy are proportional to $N$ at large $N$.

In most systems, including the networks that we study here, there are relatively few states that have high probability, and many more states with low probability; mathematically, $n(E)$ is an increasing function.  At large $N$, this competition between decreasing probability and increasing numerosity picks out a special value of $E = E^*$, which is the energy of   the ``typical'' states that we actually see; $E^*$ is  the solution to the equation
\begin{equation}
{{dS(E)}\over{dE}} = 1.
\label{fixT}
\end{equation}
For most systems,  the energy $E({\mathbf\sigma})$  has only small fluctuations around $E^*$,  $\langle (\delta E)^2 \rangle/(E^*)^2 \sim 1/N$, and in this sense most of the states that we see have the {\em same} value of log probability per degree of freedom.  But hidden in the function $S(E)$ are all the parameters describing the interactions among the $N$ degrees of freedom in the system.  At special values of these parameters,  $[d^2 S(E)/dE^2]_{E=E^*} \rightarrow 0$, and the variance of $E$ diverges as $N$ becomes large.  This is a critical point, and is mathematically equivalent to the divergence of the specific heat in an equilibrium system \cite{schnabel+al_11}.

These observations focus our attention on the ``density of states'' $n(E)$.  Rather than asking  how often we will see specific combinations of neurons  spiking while the others remain silent,  we ask how many states there are with a particular probability.  If we can estimate $n(E)$ for patterns of neural activity, then we can construct a thermodynamics for the network.

\begin{figure}[b]
\includegraphics[width=\linewidth]{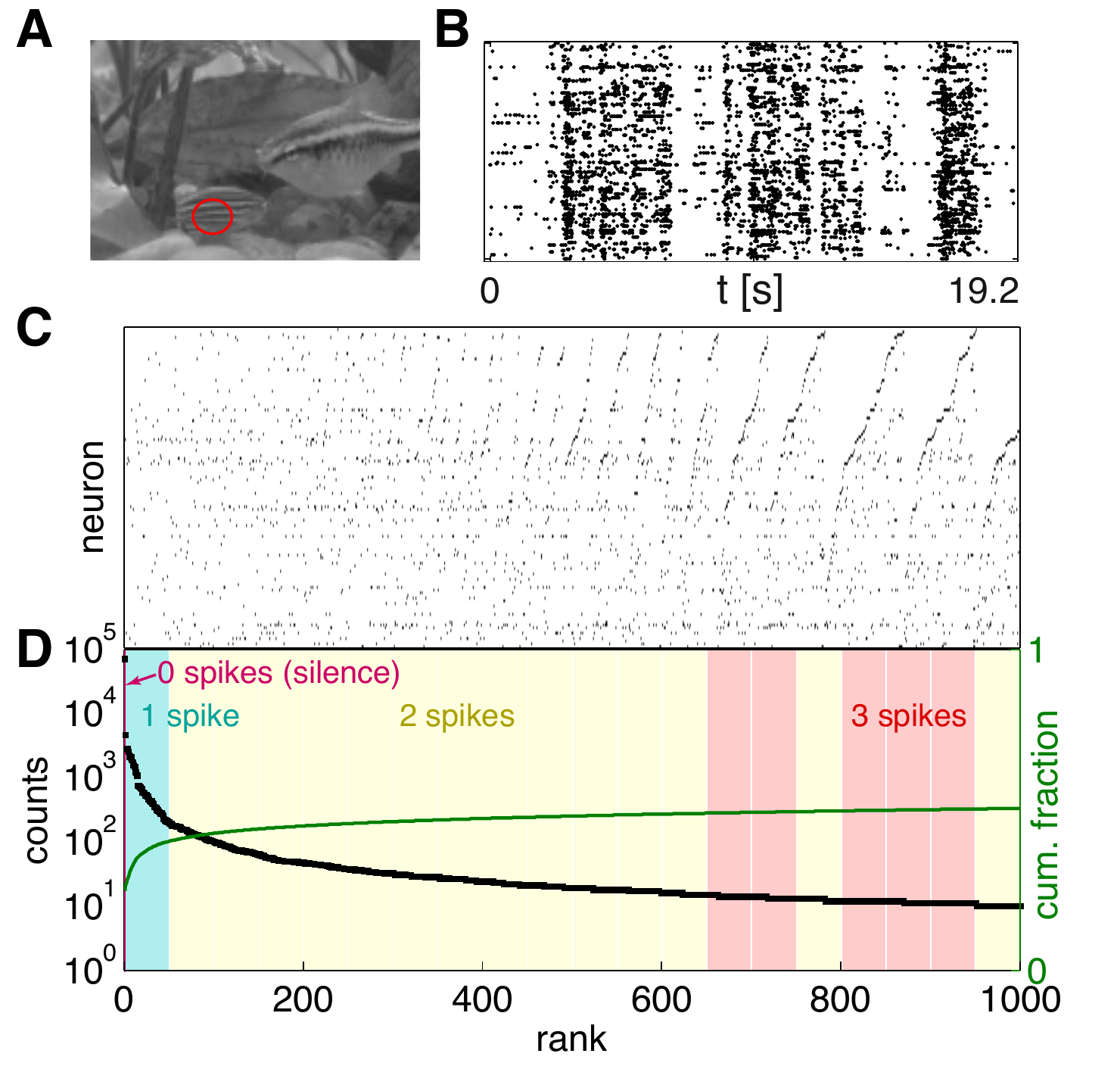}
\caption{{\bf Counting states in the response of retinal ganglion cells.} {\bf  (A)} A single frame from the naturalistic movie, red ellipse indicates the approximate extent of a receptive field center for a typical retinal ganglion cell. {\bf (B)} Responses of $N=160$ neurons to a $19.2\,{\rm s}$ naturalistic movie clip; dots indicate the times of action potentials from each neuron.  In subsequent analyses these events are discretized into binary (spike/silence) variables in time slices of $\Delta\tau = 20\,{\rm ms}$.   {\bf (C)} 1000 most common binary patterns of activity across $N=160$ neurons, in order of their frequency.  {\bf (D)} Number of occurrences of each pattern, with labels for the total number of spikes in each pattern. \label{fig1}}
\end{figure}

\section{An experimental example}

The vertebrate retina offers a unique system in which the activity of most of the neurons comprising a local circuit can be monitored simultaneously using multi--electrode array recordings. As described more fully in Ref \cite{tkacik+al_14a}, we stimulated salamander retina  with naturalistic grayscale movies of fish swimming in a tank (Fig \ref{fig1}A), while recording from 100--200 retinal ganglion cells (RGCs), the output cells of the retina that in an intact animal project to the central brain; additional experiments used artificial stimulus ensembles, as described in Appendix \ref{methods}. Sorting the raw data \cite{marre+al_12}, we could reliably identify single spikes from 160 neurons whose activity passed our quality checks and was stable for the whole $\sim$2 hour duration of the experiment; a small segment of the data is shown in Fig \ref{fig1}B. Importantly, our experiments monitored a substantial fraction of the RGCs in the area of the retina from which we record, capturing the behavior of an almost complete  local population responsible for encoding a small patch of the visual world.  The experiment collected a total of $\sim 2\times 10^6$ spikes,  and time was discretized in bins of duration $\Delta\tau = 20\,{\rm ms}$; for each neuron ${\rm i}$, $\sigma_{\rm i}=1$ in a bin denotes that the neuron emitted at least one spike, and $\sigma_{\rm i}=0$ that it was silent.

\section{Counting states}

Conceptually, estimating the function $n(E)$ and hence the entropy vs energy is easy:  we count how often each state occurs, thus estimating its probability, and then count how many states have (log) probabilities in a given range.  In Fig \ref{fig1}C and D we show the first steps in this process.  We identify the unique patterns of activity---combinations of spiking and silence across all 160 neurons---that occur in the experiment, and then count how many times each of these patterns occurs.

Even without trying to compute $S(E)$, the results of Fig \ref{fig1}D are surprising.  With $N$ neurons that can either spike or remain silent, there are $2^N$ possible states.  We know that not all these states can be visited equally often, since spikes are less common than silences, but  even taking account of this bias, and trying to capture the correlations among neurons, our best estimate of the entropy for the patterns of activity we observe is $S\sim 0.15\,{\rm bits}/{\rm neuron}$ (see below).  With $N=160$ cells, this is a total entropy of $S = 24\,{\rm bits}$, which means that the patterns of activity are spread over $2^S \sim 1.67\times 10^7$ possibilities.   This is one hundred times larger than the number of samples that we collect during our experiment.  Indeed, most of the states that we see in the full population occur only once.  But roughly one thousand states occur with sufficient frequency that we can make a reasonable estimate of their probability just by counting across the $\sim 2\,{\rm hrs}$ of the experiment.    Thus, the probability distribution $P({\mathbf \sigma})$ is extremely inhomogeneous.

To probe more deeply into the long tail of low probability events, we can construct models of the distribution of states.  As described in detail elsewhere \cite{tkacik+al_14a}, we have done this using the maximum entropy construction \cite{jaynes_57}:  we take from experiment certain average behaviors of the network, and then search for models that match these data but otherwise have as little structure as possible.   This approach works if we can show that matching a relatively small number of features produces a model which predicts many other aspects of the data.

\begin{figure}[b]
\includegraphics[width=\linewidth]{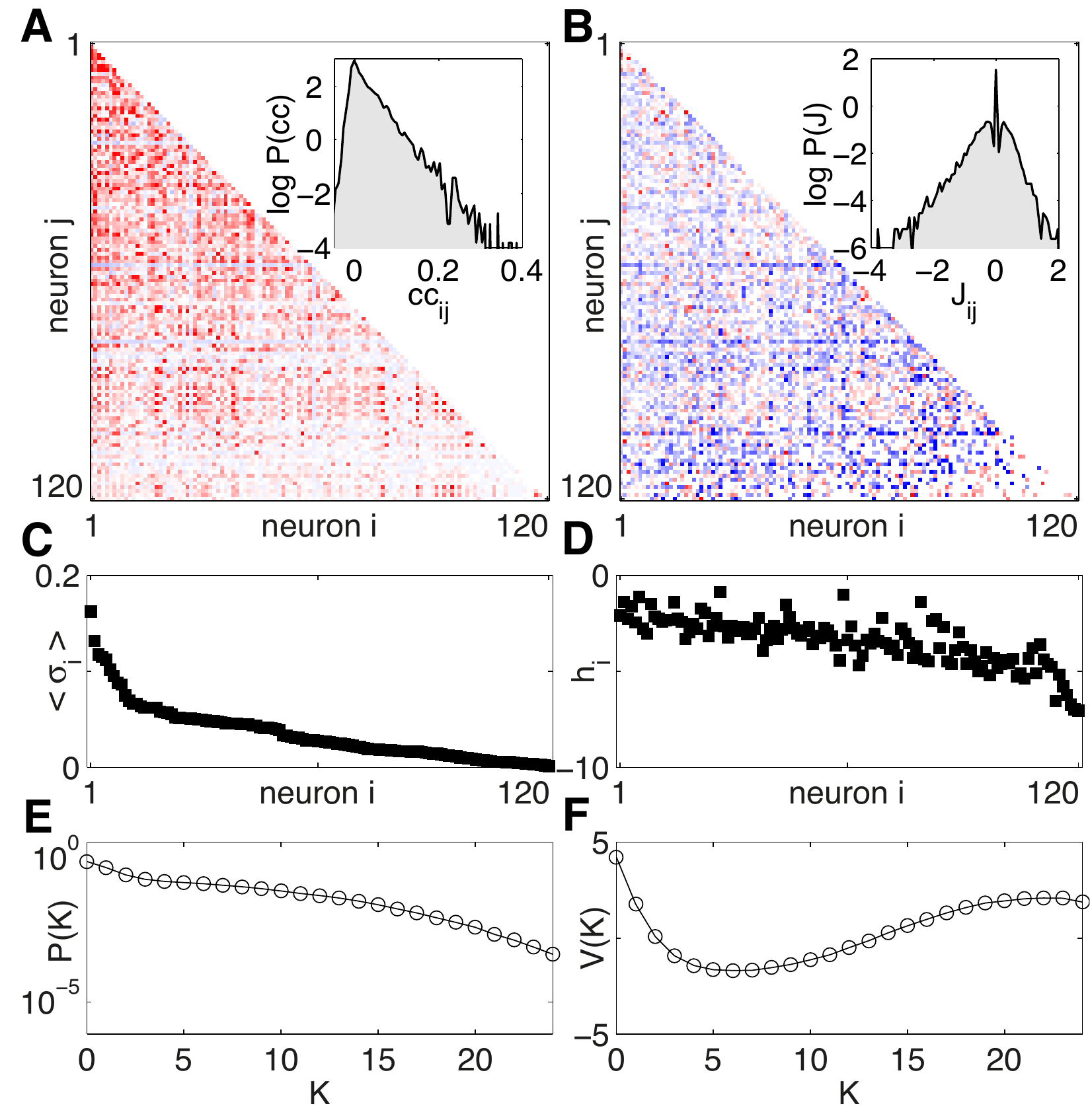}
\caption{Maximum entropy models for retinal activity in response to natural movies \cite{tkacik+al_14a}.  {\bf (A)} The correlation coefficients between pairs of neurons (red = positive, blue = negative) for a 120-neurons subnetwork. Inset shows the distribution of the correlation coefficients over the population.  {\bf (B)} The pairwise coupling matrix of the inferred model, $J_{\rm ij}$ from Eq (\ref{HKpair}). Inset shows the distribution of these pairwise couplings across all pairs $\rm ij$. {\bf (C)} The average probability of spiking per time bin for all neurons (sorted). {\bf (D)} The corresponding bias terms $h_{\rm i}$ in Eq (\ref{HKpair}).
{\bf (E)} The probability $P(K)$ that $K$ out of the $N$ neurons spike in the same time bin. {\bf (F)} The corresponding global potential $V(K)$ in Eq (\ref{HKpair}). Notice that panels A, C, and E describe the statistical properties observed for these neurons, while panels B, D, and F describe parameters of the maximum entropy model that reproduces these data within experimental errors. \label{models}}
\end{figure}

\begin{figure*}[tb]
\includegraphics[width=\linewidth]{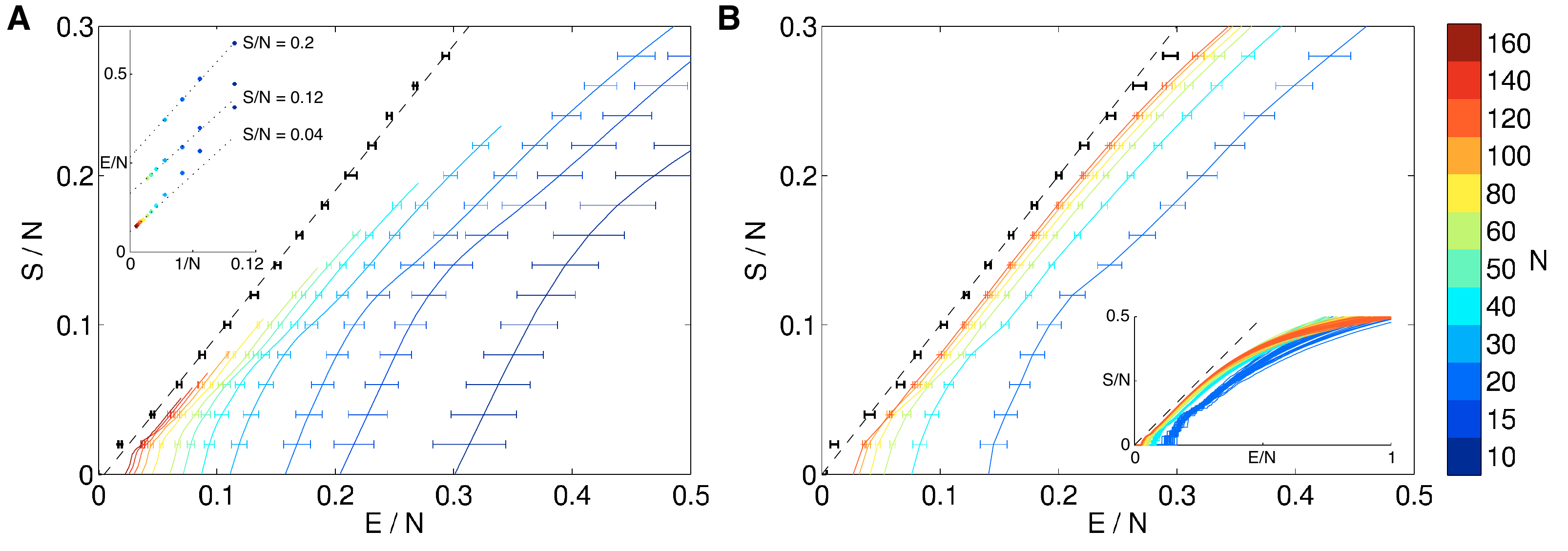} 
\caption{{\bf Entropy vs. energy.  (A)} Computed directly from the data using Eqs (\ref{cumNE}, \ref{SfromN}).  Different colors show results for different numbers of neurons; in each case we choose 1000 groups of size $N$ at random, and points are means with standard deviations over groups. Inset shows extrapolations of the energy per neuron at fixed energy per neuron, summarized as black points with error bars in the main figure.  Dashed line is the best linear fit to the extrapolated points, $S/N = (0.974 \pm 0.021) (E/N) + (-0.005 \pm 0.003)$. {\bf (B)} Computed from the maximum entropy models.   We choose $N = 20,\,  40,\, \cdots ,\, 120$ neurons out of the population from which we record, and for each of these subnetworks we construct maximum entropy models as in Eq (\ref{HKpair}); for details of the entropy calculation see Appendix \ref{app:DoS1}.   Inset shows results for many subnetworks, and the main figure shows means and standard deviations across the different groups of $N$ neurons.  Black points with error bars are an extrapolation to infinite $N$, as in {\bf (A)},  and the dashed line is $S = E$.  \label{fig3}}
\end{figure*}

The maximum entropy approach to describing activity in networks of neurons has been explored, in several different systems, for nearly a decade \cite{schneidman+al_06,tkacik+al_06,shlens+al_06,tang+al_08,tkacik+al_09,shlens+al_09,ohiorhenuan+al_10,ganmor+al_11,simplest,sdme}, and there have been parallel efforts to use this approach in other biological systems \cite{lezon+al_06,tkacik_07,bialek+ranganathan_07,seno+al_08,weigt+al_09,stephens+bialek_10,mora+al_10,marks+al_11,sulkowska+al_12,bialek+al_13,mukherjee+al_13,bialek+al_14,santolini+al_14}.  Recently we have used the maximum entropy method to build models for the activity of up to $N=120$ neurons in the experiments described above  \cite{tkacik+al_14a}.   As summarized in Fig \ref{models}, we take from experiment the mean probability of each neuron generating a spike ($\langle \sigma_{\rm i}\rangle$), the  correlations between spiking in pairs of neurons ($\langle \sigma_{\rm i}\sigma_{\rm j}\rangle$), and the probability that $K$ out of the $N$ neurons spike in the same small window of time [$P(K)$].  Mathematically, the maximum entropy models consistent with these data have the form
\begin{eqnarray}
P(\{\sigma_{\rm i}\}) &=& {1\over Z} \exp\left[ - E(\{\sigma_{\rm i}\})\right]\label{PH} \\
E (\{\sigma_{\rm i}\})&=& - \sum_{{\rm i}=1}^N h_{\rm i} \sigma_{\rm i} - {1\over 2} \sum_{{\rm i}, {\rm j} =1}^N J_{\rm ij} \sigma_{\rm i} \sigma_{\rm j}
- V \left(K \right) ,  \label{HKpair}
\end{eqnarray}
where $K = \sum_{{\rm i}=1}^N \sigma_{\rm i}$ counts the number of neurons that spike simultaneously, and $Z$ is set to insure normalization.  All of the parameters $\{h_{\rm i},\, J_{\rm ij},\, V(K)\}$ are determined by the measured averages $\{\langle \sigma_{\rm i}\rangle,\, \langle \sigma_{\rm i}\sigma_{\rm j}\rangle,\, P(K)\}$.

This model accurately predicts the correlations among triplets of neurons (Fig 7 in \cite{tkacik+al_14a}), and  the probability of spiking in individual neurons depends on the pattern of activity in the rest of the population as the model predicts (Fig 9 in \cite{tkacik+al_14a}).  One can even predict the time dependent response of single cells from the behavior of the population, without reference to the visual stimulus (Fig 15 in \cite{tkacik+al_14a}).  Most important for our present discussion, we have checked that the distribution of the energy $E (\{\sigma_{\rm i}\})$ across the patterns of activity that occur in the data agrees with the distribution predicted by the model, deep into the tail of patterns that occur only once in the two hour long experiment (Fig 8 in \cite{tkacik+al_14a}).  This distribution is very closely related to the plot of entropy vs energy that we would like to construct, and so the agreement with experiment gives us confidence.

The direct counting of states (Fig \ref{fig1}) and the maximum entropy models (Fig \ref{models}) give us two complementary ways of estimating the function $n(E)$ and hence the entropy vs energy in the same data set.  Results are shown in Fig \ref{fig3}, with some technical issues discussed in Appendix \ref{app:DoS1}.

As emphasized above and in Appendix \ref{thermoP}, the plot of entropy vs energy contains all of the thermodynamic behavior of a system, and this has a meaning for any probability distribution, even if we are not considering a system at thermal equilibrium.  Thus, Fig \ref{fig3} is as close as we can get to constructing the thermodynamics of this network.  With the direct counting of states we see less and less of the plot at larger $N$, but   the part we can see is approaching a limit as $N \rightarrow \infty$, and this is confirmed by the results from the maximum entropy models.  This by itself is a significant result.  If we write down a model like Eq (\ref{HKpair}), then in a purely theoretical discussion we can scale the couplings between neurons $J_{\rm ij}$ with $N$ to guarantee the existence of a thermodynamic limit \cite{amit+al_87}, but  with $J_{\rm ij}$ constructed from real data, as we do here, we can't impose this scaling ourselves---either it emerges from the data or it doesn't. We can make the emergence of the thermodynamic limit more precise by noting that, at a fixed value of $S/N$, the value of $E/N$ extrapolates to a well defined limit in a plot vs. $1/N$, as  in the inset to Fig \ref{fig3}A.  The results of this extrapolation from the real data are strikingly simple:  the entropy is equal to the energy, within (small) error bars.   
.

\section{Interpreting $S(E)$}

What thermodynamic behavior is predicted by Fig \ref{fig3}?  If the plot of entropy vs energy is a straight line with unit slope, then Eq (\ref{fixT}) is solved not by one single value of $E$ but by a whole range.  Not only do we have $d^2 S/dE^2 = 0$, as at an ordinary critical point, but all higher order derivatives also are zero.  Thus, the results in Fig \ref{fig3}  suggest that the joint distribution of   activity across neurons in this network is poised at a very unusual critical point.  

Figure \ref{fig3} is telling us that the tradeoff between the probability and numerosity of states in the network takes a very special form.  We expect that states of lower probability (e.g., those in which more cells spike) are more numerous (because there are more ways to arrange $K$ spikes among $N$ cells as $K$ increases from very low values).  But the usual result is that this tradeoff---which is precisely the tradeoff between energy and entropy in thermodynamics---selects  ``typical'' states that all have roughly the same probability.  The statement that $S(E) =E$, as suggested  in Fig \ref{fig3}, is the statement that states which are ten times less probable are {\em exactly} ten times more numerous, and so there is no typical value of the probability.  This balancing of probability and numerosity extends over a range of at least $0.2$ along the $E/N$ axis, corresponding to a factor of  $\sim e^{(0.2)N}\sim 10^{10}$ in probability for networks of $N=120$ neurons.

\section{Heat capacities}

The divergence of the specific heat is one of the classical signs of criticality in equilibrium thermodynamic systems.  Although the neurons obviously are not an equilibrium system, the model probability distribution in Eqs (\ref{PH}, \ref{HKpair}) is mathematically identical to the Boltzmann distribution for a system in equilibrium at temperature $k_BT =1$.  Thus we can take this model seriously as a statistical mechanics problem, and compute the specific heat in the usual way; for details see Appendix \ref{app:DoS1}.  Furthermore, we can  change the effective temperature by considering a one parameter family of models,
\begin{equation}
P(\{\sigma_{\rm i}\}; T) = {1\over {Z(T)}} \exp\left[ -{1\over T}E (\{\sigma_{\rm i}\})\right] ,
\end{equation}
with $E (\{\sigma_{\rm i}\})$ as before in Eq (\ref{HKpair}).  The goal is to see whether there is anything special about the value $T=1$ that describes the real system.

\begin{figure}[tb]
\includegraphics[width=\linewidth]{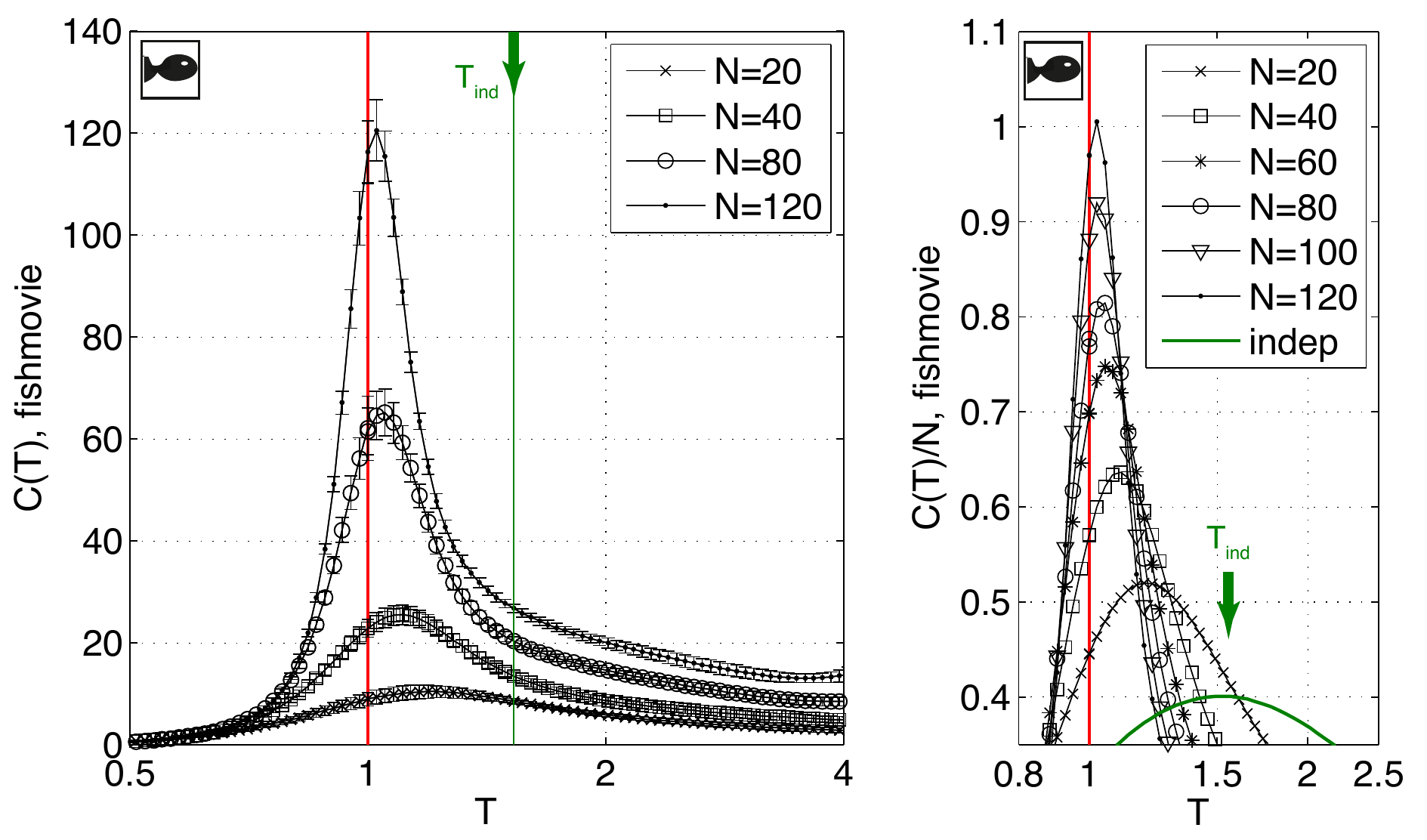}
\caption{{\bf Heat capacity in maximum entropy models of neurons responding to naturalistic stimuli.} {\bf (A)} Heat capacity, $C(T)$, computed for subnetworks of  $N=20,\, 40,\, 80,\, 120$, neurons.  Points are averages, and error bars are standard deviations, across 30 choices of subnetwork at each $N$.   For details of the computations see Appendix \ref{app:DoS1}.  {\bf (B)} Zoom--in of the peak of $C(T)$, shown as an intensive quantity, $C(T)/N$. The peak moves towards $T=1$ with increasing $N$, and the growth of $C(T)$ near $T=1$ is faster than linear with $N$. Green arrow points to results for a model of independent neurons whose spike probabilities (firing rates) match the data exactly; the heat capacity is exactly extensive.\label{Cv}}
\end{figure}

Results for the heat capacity of our model vs $T$ are shown in Fig \ref{Cv}.   There is a dramatic peak in this plot, and as  we look at larger and larger groups of neurons, the peak grows and moves closer to $T=1$, which is the model of the actual network.  Importantly, the heat capacity grows even when we normalize by $N$, so that the specific heat, or heat capacity per neuron, is growing with $N$, as expected at a critical point.

Temperature and energy may seem like foreign concepts in the context of real neurons, so the results of Fig \ref{Cv} require some interpretation.  We have studied one particular network, that is described by some set of parameters.  Are  the parameters that describe the real network in any sense special?  Changing ``temperature'' allows us to probe along one axis in parameter space, and the fact that we observe a peak in the specific heat means that the real network in poised in parameter space very close to a maximum in the variance of log(probability), which we can think of as the dynamic range of surprise that can be represented by the network.

\begin{figure}[b]
\includegraphics[width=\linewidth]{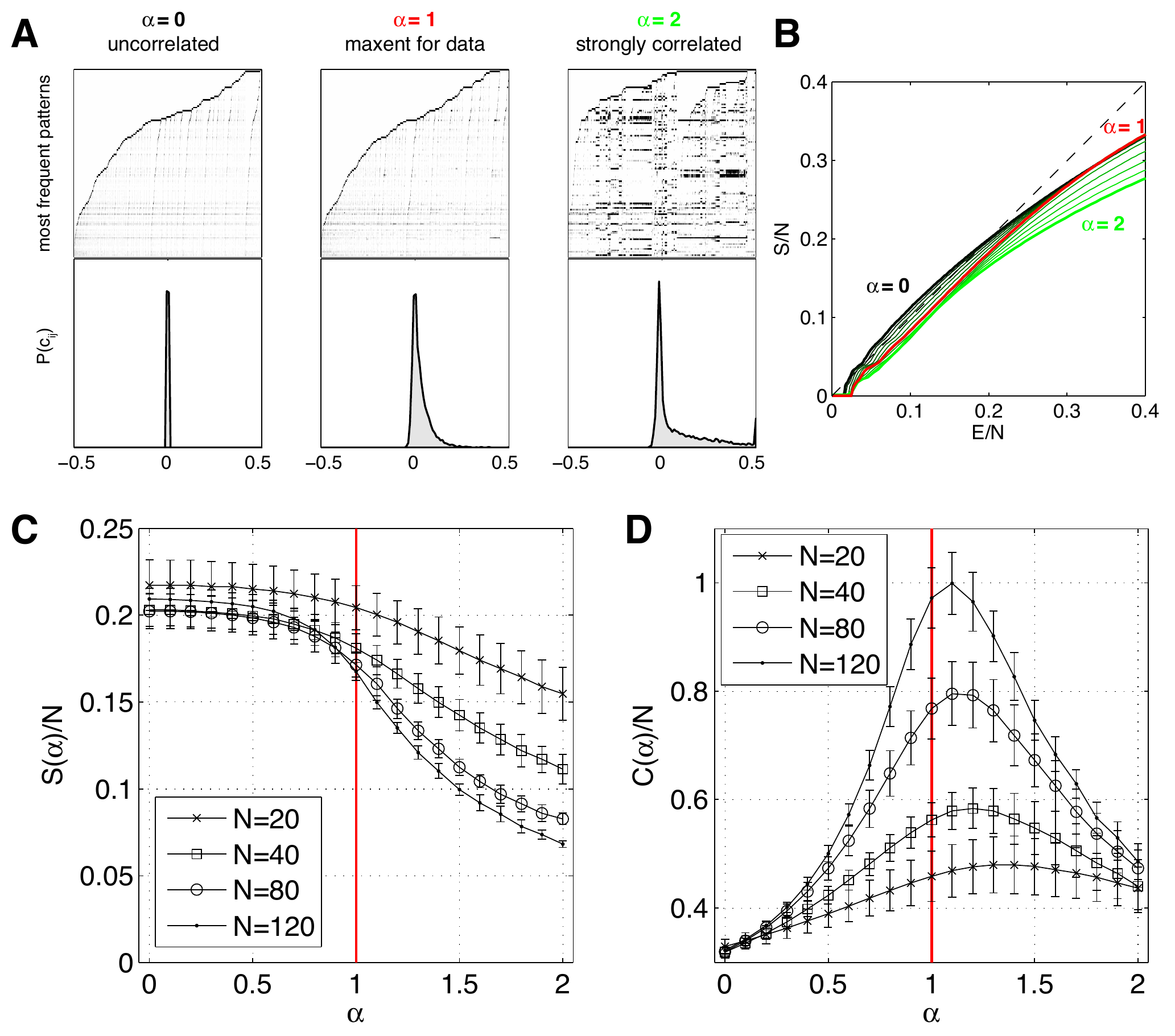}
\caption{{\bf Changing correlations at fixed spike rates.} {\bf (A)} Three maximum entropy models for a 120-neuron network, where correlations have been eliminated (left, $\alpha=0$), left at the strength found in data (middle, $\alpha=1$), or  scaled up (right, $\alpha=2$). The first row shows 10000 most frequent patterns (black=spike, white=silence) in each model. The second row shows the distribution of pairwise correlation coefficients. {\bf (B)} Entropy vs. energy for the networks in A. {\bf (C)} Entropy per neuron as a function of $\alpha$, for different subnetwork sizes $N$.  {\bf (D)} Heat capacity per neuron exhibits a peak close to $\alpha=1$.  Error bars are  standard deviations over 10 subnetworks for each $N$ and $\alpha$.\label{alpha_fig}}
\end{figure}

The temperature is only one axis in parameter space, and along this direction there are variations in both the correlations among neurons and their mean spike rates.  As an alternative, we can consider a family of models in which the strength of correlations changes but spike rates are fixed.  Such a family is given by 
\begin{eqnarray}
P(\{\sigma_{\rm i}\}; \alpha) &=& {1\over {Z(\alpha)}} \exp\left[ -E_\alpha (\{\sigma_{\rm i}\})\right]\label{PHalpha} \\
E_\alpha (\{\sigma_{\rm i}\})&=& - \sum_{{\rm i}=1}^N h_{\rm i}' (\alpha )\sigma_{\rm i} 
\nonumber\\
&&\,\,\,\,\,\,\,\,\,- \alpha\left[ {1\over 2} \sum_{{\rm i}, {\rm j} =1}^N J_{\rm ij} \sigma_{\rm i} \sigma_{\rm j}
+ V \left(K \right) \right],  \label{HKpairalpha}
\end{eqnarray}
where changing $\alpha$ changes the strength of correlations, and we adjust all the $h_{\rm i}' (\alpha )$ to hold mean spike rates fixed at their observed values.   The behavior of this family of models is explored in Fig \ref{alpha_fig}.

At $\alpha=0$, the model describes a population of independent neurons, so that the distribution of correlation coefficients across the populations consists of a singular peak at $C=0$ (Fig \ref{alpha_fig}A, left panel).  As $\alpha$ is increased beyond the strength of correlations present in the data ($\alpha=1$, Fig \ref{alpha_fig}A, middle panel), the distribution of correlations broadens such that at $\alpha=2$ some pairs are very strongly correlated (Fig \ref{alpha_fig}A, right panel). This is reflected in a distribution of frequent joint activity states that seem to cluster around a small number of prototypical patterns, much as   in the Hopfield model of associative memories \cite{hopfield_82,amit_89}. The entropy vs energy plot for the corresponding network, shown in Fig 5B, singles out the ensemble at $\alpha=1$ as the one whose density of states aligns most closely with the diagonal: going towards independence (smaller $\alpha$) gives rise to a concave bump at low energies, while $\alpha>1$ ensembles deviate away from the equality line more at high energies. Interestingly, the drop in entropy from independence to $\alpha=1$ is modest, consistent with our previous reports \cite{tkacik+al_14a}, and the entropy decreases substantially only at high $\alpha$ and large $N$, as shown in Fig \ref{alpha_fig}C.  Corresponding to the variations in $S(E)$ shown in Fig \ref{alpha_fig}B, we see in Fig \ref{alpha_fig}D that there is a peak in the specific heat of the model ensemble near $\alpha=1$.  As we look at larger and larger networks, this peak rises  and moves toward $\alpha = 1$, which describes the real system.

\section{Couldn't it just be ... ?}

In equilibrium thermodynamics, poising a system at a critical point involves careful adjustment of temperature, pressure, and other parameters.  Finding that the retina seems to have poised itself near criticality should thus be treated with some skepticism.  Here we consider some of the pitfalls that might mislead us into thinking that the system is critical when it is not; see also Appendix \ref{app:alt}.

Part of our analysis is based on the use of maximum entropy models, and one could worry that the inference of these models is unreliable for finite data sets  \cite{mastromatteo+marsili_11,marsili+al_13}.  Expanding on the discussion in Ref \cite{tkacik+al_14a}, we can show that signatures of criticality, such as the peak in specific heat, are not the result of over fitting to a limited data set.  As an example, we see a clear peak in the specific heat when we learn models for $N=100$ neurons from even one--tenth of our data,  and the variance across fractions of the data is only a few percent (Appendix \ref{limited_data}, Fig \ref{Cv_finitesize}).

While the inference of maximum entropy models might be accurate, some less interesting models might mimic the signatures of criticality \cite{1overF}.  In particular, it has been suggested that independent neurons with a broad distribution of spike rates could generate a distribution of $N$--neuron activity patterns $\{\sigma_{\rm i}\}$ that mimics some aspects of critical behavior \cite{bas}.  But in an independent model built from the actual spike rates of the neurons, the probability of seeing the same state twice would be less than one part in a billion,  dramatically inconsistent with the measured $P_c \sim 0.04$. Such independent models also cannot account for the faster than linear growth of the heat capacity with $N$ (Fig \ref{Cv}), which is an essential feature of the data and its support for criticality.

In  maximum entropy models, the  probability distribution over patterns of neural activity is described in terms of ``interactions'' between neurons, such as the terms $J_{\rm ij}$ in Eq (\ref{HKpair}); an alternative view is that the correlations  result from the response of the neurons to fluctuating external signals.  Testing this idea has a difficulty that has nothing to do with neurons.  Going back to equilibrium statistical mechanics, many models in which spins (or other degrees of freedom) interact with one another are mathematically equivalent to a collection of spins responding independently to a fluctuating field (Appendix  \ref{SNM}).  Thus, correlations are almost always interpretable as independent responses to unmeasured fluctuations, and for neurons there are many possibilities, including internal states of neurons, that would not fall into the usual classification of network effects vs common input.    

We can exclude the visual stimulus alone as the explanation of what we see, since similar behavior---e.g., the growth of a peak in specific heat with increasing $N$---appears in response to random checkerboard movies or even full field flicker (Appendix \ref{cond_ind}, Fig \ref{compare_ens}).  Our analysis focused on the total distribution over activity states, not distinguishing between the correlations due to a common stimulus input and those due to intrinsic circuit interactions. While a network of conditionally independent neurons might give rise to signatures of criticality as observed here, such models are incompatible with our data. We can construct conditionally independent neurons by shuffling the repeated presentations of the same movie, and in this surrogate data the probability of the whole network remaining silent changes slightly, but this change is ten times larger than the standard error in the measurement.  Conditional independence thus is excluded, based just on this one statistic, with a likelihood ratio of more than ten thousand  (Appendix \ref{cond_ind}, Fig \ref{silence}).  Apart from these arguments,  experiments show that electrically evoked spikes in retinal ganglion cells feed back into the retinal circuitry via electrical synapses and modulate the response of other ganglion cells to visual input \cite{asari+meister_12}.  This is a direct demonstration that retinal ganglion cells are not conditionally independent encoders of their visual input, as are the more common measurements of ``noise correlations'' between ganglion cells \cite{mastronarde_83,brivanlou+al_98,field+chichilnisky_07}.

In interacting models that are equivalent to a collection of independent spins responding to a fluctuating field, criticality usually requires that the distribution of fluctuations be very special, e.g. with the variance tuned to a particular value.  In this sense, saying that correlations result from fluctuating inputs wouldn't really ``explain'' our observations.    Recently, it has been suggested that sufficiently broad distributions of fluctuations might lead more generically to critical phenomenology \cite{schwab+al_13}, but this argument requires that the number of neurons be much larger than the number of independently fluctuating fields.   If  the relevant fields are generated by retinal neurons that provide input to the ganglion cells, then  changing the effective dimensionality of the stimulus---from full field flicker to natural movies to random checkerboards---would change the number of effective fields, and at one extreme this number is comparable to or larger than the number of neurons from which we record.  Nonetheless we see near--critical behavior in all these cases.  It is also not clear how such models with fluctuating fields would explain the fact that networks with slightly stronger or weaker correlations among neurons both deviate from criticality (Fig \ref{alpha_fig}).

The evidence for criticality that we find in the analysis of 160 neurons is consistent with  extrapolations from the analysis of smaller populations under similar conditions \cite{tkacik+al_06,tkacik+al_09}.  The basis for those predictions was the assumption that the smaller populations were typical of the larger one, i.e. that the  spike probabilities and pairwise correlations that we could observe were drawn from the same distribution as in the full system, and that these distributions (rather than the detailed structure of the correlation matrix, for example) were sufficient to determine the thermodynamic behavior \cite{castellana+bialek_13}.      Hints of criticality also are observable in vastly simpler models, which capture the distribution of summed activity in the network but ignore the identity of individual neurons \cite{simplest}.

\section{Discussion}

The traditional formulation of the neural coding problem makes an analogy to a dictionary, asking for the ``meaning'' of each neural response in terms of events in the outside world \cite{spikes}.  But before we can build a dictionary, we need to know the lexicon, and for large populations of neurons this already is a difficult problem:  what is the set of responses that the network actually uses?   With 160 neurons, the number of possible responses is larger than the number of words in the vocabulary of a well educated English speaker, and is more comparable to the number of possible short phrases or sentences.  In passing from letters to words to sentences, we encounter many features of language that leave an imprint on the probability distribution:  the joint distribution of letters in words embodies spelling rules \cite{stephens+bialek_10}, while the joint distribution of words in sentences  encodes aspects of grammar \cite{pereira_00} and semantic categories \cite{pereira+al_93}.  By analogy,   the joint distribution of activity among neurons  should reveal structures that have biological significance.

The small patch of the retina that we consider has several types of ganglion cells, and indeed  no two cells have truly identical input/output characteristics, even in response to spatial uniform stimuli \cite{schneidman+al_03}.  Nonetheless,   groups of twenty or more cells have collective properties that vary very little across different groups of neurons.  If we ask not about the probability of each detailed pattern of spiking and silence, but rather count how many combinations have a given probability, this relationship is highly reproducible from group to group, and  simplifies as we look at larger and larger groups.  

The relationship between probability and numerosity of states is mathematically identical to the relationship between energy and entropy in statistical physics.  The fact that this relationship simplifies as we look at larger groups of neurons suggests that we are seeing signs of a thermodynamic limit.   As  emphasized above, the existence of such a limit in a real network is not guaranteed.

If we can identify the thermodynamic limit, we can try to place the network in a  phase diagram of possible networks.    Crucial to our analysis is that the critical surfaces which separate different phases have signatures that are detectable even if we don't know the nature of the phases themselves.  In particular, criticality is associated with a finely balanced tradeoff between probability and numerosity:  states that are a factor $F$ times less probable are also a factor $F$ times more numerous.  At conventional critical points, this balancing occurs only in a small neighborhood of the typical probability, but in the network of retinal ganglion cells  we see nearly perfect balancing across a wide range of probabilities (Fig \ref{fig3}).  If we construct model networks that have slightly stronger or weaker correlations among pairs of neurons, then in both cases we see a breakdown of this balance (Fig \ref{alpha_fig}).  

The strength of correlations depends on the structure of visual inputs, on the connectivity of the neural circuit, and on the state of adaptation in the system.  The fact that we see signatures of criticality in response to very different visual inputs (Appendix \ref{cond_ind}), but that these signatures break down in model networks with stronger or weaker correlations, strongly suggests that adaptation is tuning the system toward a critical state.  This predicts that a sudden change of visual input statistics will drive the network to a non--critical state, and that during the course of adaptation the distribution of activity will relax back to the critical surface.  This can be tested directly.

Does criticality have functional consequences?  The crucial signature of criticality that we have seen in the data is the extreme inhomogeneity of the probability distribution over states.  This allows the system to construct an instantaneously readable code for events in the world that have a large dynamic range of likelihoods or surprise, and this may be well suited to the challenges of the natural environment; it is not, however, an efficient code in the usual sense.  Systems near critical points are maximally responsive to certain external signals, and this sensitivity may be functionally useful.  Most of the systems that exhibit criticality in the thermodynamic sense also exhibit a wide range of time scales in their dynamics, so that criticality may provide a general strategy for neural systems to bridge the gap between the microscopic time scale of spikes and the macroscopic time scales of behavior.   Critical states are extremal in all these different senses, and more;  in the examples we understand, these different features  also are not separable, so it may be difficult to decide which is relevant for the organism.   

In addition to the the network of neurons studied here,  closely related signatures of criticality have been detected in ensembles of amino acid sequences for protein families \cite{mora+al_10}, in flocks of birds \cite{bialek+al_14} and swarms of insects \cite{attanasi+al_13}, and in the network of genes controlling morphogenesis in the early fly embryo \cite{krotov+al_14};  there is also evidence that cell membranes have lipid compositions tuned to a true thermodynamic critical point \cite{honerkamp-smith+al_09}.  Different, dynamical notions of criticality have been explored in neural \cite{beggs+plenz_03,friedman+al_12} and genetic \cite{nykter+al_08,balleza+al_08} networks, as well as in the active mechanics of the inner ear \cite{eguiluz+al_00,calamet+al_00,ospeck+al_01}.  This convergence may hint at a general principle, but there is considerable room for skepticism \cite{mora+bialek_11,beggs+timme_12}.   A new generation of experiments on larger populations of neurons should allow for decisive tests of our ideas about collective behavior in these networks, including the possibility of criticality.

\begin{acknowledgements}
We thank A Cavagna, I Giardina, M Ioffe, SCF van Opheusden, SE Palmer,  E Schneidman, DJ Schwab, and AM Walczak  for helpful discussions.   Research supported in part by NSF Grants  PHY--1305525 and CCF--0939370,  by NIH  Grant R01 EY14196, and by Austrian Science Foundation Grant FWF P25651.  Additional support provided by the Fannie and John Hertz Foundation,  by  the Swartz Foundation, and by the WM Keck Foundation.
\end{acknowledgements}

\appendix

\section{Thermodynamics and probability distributions}
\label{thermoP}

The fundamental variables of thermodynamics are energy, temperature, and entropy.  For the states taken on by a network of neurons, energy and temperature are meaningless, so it is difficult to see how we can construct a thermodynamics for these systems.  But in statistical mechanics, all thermodynamic quantities are derivable from the Boltzmann distribution, the probability that the system will be found in any particular state.  Thus, all thermodynamic statements can be seen as statements about this underlying probability distribution, and in this sense we should be able to construct thermodynamics for {\em any} probability distribution that describes a large number of variables.

The idea that all probability distributions over $N$ variables have an associated thermodynamics in the $N\rightarrow\infty$ limit is powerful, but perhaps not so widely used.  This connection is well studied by mathematical physicists \cite{ruelle}, and has been a guide to the analysis of experiments on dynamical systems \cite{halsey+al_86,feigenbaum+al_86}.    We have used these ideas to construct a thermodynamics of natural images \cite{stephens+al_13}, and have emphasized the connection of thermodynamic criticality to Zipf's law \cite{mora+bialek_11}.  Here we give a somewhat pedagogical discussion, in the hope of making the results accessible to a broader audience.

We start by recalling that, for a system in thermal equilibrium at temperature $T$, the probability of finding the system in state $s$ is given by
\begin{equation}
P_s = {1\over Z}e^{-E_s/k_B T} ,
\end{equation}
where $E_s$ is the energy of the state, and Boltzmann's constant $k_B$ converts between conventional units of temperature and energy.  The partition function $Z$ serves to normalize the distribution, which requires
\begin{equation}
Z = \sum_s e^{-E_s/k_B T} ,
\end{equation}
but in fact this normalization constant encodes many physical quantities.  The logarithm of the partition function is proportional to the free energy of the system, the derivative of the free energy with respect to the volume occupied by the system is the pressure, the derivative with respect to the strength of an applied magnetic field is the magnetization, and so on.  

The ``state'' of a system is defined by the joint configuration of all its parts.  Thus in a classical gas or liquid, $s$ is defined by the positions and velocities of all the constituent atoms.  Different gases or liquids differ not because these variable are different, but because the energy $E_s$ is a different function of these $N$ underlying variables.  But thermodynamics doesn't make reference to all these details.  Which aspects of the underlying microscopic rules actually matter for predicting the free energy and its derivatives?

We can write the sum over all states as a sum first over states that have the same energy, and then a sum over energies.  We do this by introducing an integral over a delta function into the sum:
\begin{eqnarray}
Z &=& \sum_s e^{-E_s/k_B T} \nonumber\\
&=& \sum_s \left[\int dE \,\delta (E- E_s)\right] e^{-E_s/k_B T} \\
&=&\int dE \sum_s \delta (E- E_s)e^{-E_s/k_B T}\\
&=& \int dE\, e^{-E/k_B T} \left[ \sum_s \delta (E- E_s)\right] .
\label{nextstep}
\end{eqnarray}
We see that the way in which the energy depends on each state appears only in the brackets, a function $n(E)$ that counts how many states have a particular energy.  

Looking ahead to the analysis of real data, it will be convenient to rearrange Eq (\ref{nextstep}) slightly.  Instead of counting the number of states that have energy $E$, we can count the number of states with energy less than $E$:
\begin{equation}
{\cal N}(E) = \sum_s \Theta\left(E -  E_s\right) ,\label{cumNE}
\end{equation}
where the step function is defined by
\begin{eqnarray}
\Theta\left(x>0\right) &=& 1\\
\Theta\left(x< 0\right) &=& 0.
\end{eqnarray}
But the step function is the integral of the delta function, which means that we can integrate by parts in Eq (\ref{nextstep}) to give
\begin{equation}
Z = {1\over {k_B T}} \int dE\, e^{-E/k_B T} {\cal N} (E).
\label{ZfromN}
\end{equation}

If we think about $N$ variables, each of which can take on only two states, the total number of states is $2^N$.  More generally, we expect that the number of possible states in a system with $N$ variables is exponentially large, so it is natural to think not about the number of states ${\cal N}(E)$ but about its logarithm, 
\begin{equation}
S(E) = \ln {\cal N}(E),\label{SfromN}
\end{equation}
which is called the entropy. 

As a technical aside, we can define the entropy either in terms of the number of states with energy close to $E$, what we have called $n(E)$ in the main text, or we can use the number of states with energy less than $E$, what we have called ${\cal N}(E)$.  In the limit that the number of degrees of freedom in the system become large, there is no difference in the resulting estimate of the entropy per degree of freedom,  because the number of states is growing exponentially fast with the energy, so that the vast majority of states with energy less than $E$ also have energy very close to $E$.  When it comes time to analyze experimental data, however, using ${\cal N}(E)$ allows us to avoid making bins along the $E$ axis.

Substituting from Eq (\ref{SfromN}) into Eq (\ref{ZfromN}), the partition function can be written as an integral determined only the function $S(E)$, entropy vs energy:
\begin{equation}
Z = {1\over {k_B T}} \int dE\, \exp\left[ -{E\over {k_B T}} +  S(E)\right] .
\label{Z-1}
\end{equation}

One of the key ideas in thermodynamics is that certain variables are ``extensive,'' that is proportional to the number of particles or variables in the system, while other variables are ``intensive,'' independent of the system size.  Temperature is an intensive variable, energy and entropy are extensive variables.  It is then natural to think about the energy per particle $\epsilon = E/N$, and the entropy per particle, $S(E)/N = s(\epsilon )$.  In the limit of large $N$, we expect $s(\epsilon )$ to become a smooth function.  Substituting into Eq (\ref{Z-1}), the partition function can be written as
\begin{eqnarray}
Z &=& {N\over {k_B T}} \int d\epsilon \, e^{ -N f(\epsilon )/k_B T}
\label{Zf}\\
f(\epsilon ) &=& \epsilon -k_B T  s(\epsilon ) .
\label{fdef}
\end{eqnarray}
We note that $f(\epsilon )$ is the difference between energy and entropy, scaled by the temperature, and is called the free energy.

Whenever we have an integral of the form in Eq (\ref{Zf}), at large $N$ we expect that it will be dominated by values of $\epsilon$ close to the minimum of $f(\epsilon)$.  This minimum $\epsilon_*$ is the solution to the equation
\begin{equation}
{{df(\epsilon )}\over {d\epsilon}} = 0 \Rightarrow {1\over {k_B T}} = {{ds(\epsilon )}\over {d\epsilon}} ,
\label{Tdef}
\end{equation}
which we can also think of as defining the temperature. Notice that $T$ being positive requires that the system have $ds(\epsilon) /d\epsilon >0$, which means there are more states with higher energies.

If we expand $f(\epsilon )$ in the neighborhood of $\epsilon_*$, we have
\begin{equation}
f(\epsilon ) = f(\epsilon_* ) - {{k_B T}\over 2} {{d^2s(\epsilon )}\over {d\epsilon^2}} {\bigg |}_{\epsilon_*} (\epsilon - \epsilon_* )^2 + \cdots ,
\label{f_expand}
\end{equation}
which gives 
\begin{widetext}
\begin{equation}
Z = {N\over {k_B T}}  e^{- f(\epsilon_* )/k_B T} \int d\epsilon \, \exp\left( - {N\over {2 }} \left[ - {{d^2s(\epsilon )}\over {d\epsilon^2}}{\bigg |}_{\epsilon_*} \right] (\epsilon - \epsilon_* )^2 + \cdots \right) .
\label{Zsaddle}
\end{equation}
\end{widetext}
This looks as if the energy per particle is drawn from a Gaussian distribution, with mean $\epsilon_*$ and variance 
\begin{equation}
\langle (\delta \epsilon)^2\rangle =  {1\over N}  \left[-  {{d^2s(\epsilon )}\over {d\epsilon^2}}{\bigg |}_{\epsilon_*}  \right]^{-1} ,
\label{var_eps}
\end{equation}
and indeed this can be shown more directly from the Boltzmann distribution.

With the interpretation of $\epsilon_*$ as the mean energy per particle, we can use Eq (\ref{Tdef}) to calculate how this energy changes when we change the temperature, and we find
\begin{equation}
{{d\epsilon_*}\over{dT}} = {1\over{k_B T^2}}  \left[-  {{d^2s(\epsilon )}\over {d\epsilon^2}}{\bigg |}_{\epsilon_*}  \right]^{-1} .
\label{C_and_s}
\end{equation}
The change in energy with temperature is called the heat capacity $C$, and when we normalize per particle it is referred to as the specific heat.  Combining Eqs (\ref{var_eps}) and (\ref{C_and_s}), we see that the specific heat $C/N$ is connected to the variance in energies,
\begin{equation}
\langle (\delta \epsilon)^2\rangle = k_B T^2 {C\over N}  .
\label{CandVarE}
\end{equation}
This relationship also can be proven without resorting to the approximation in Eq (\ref{f_expand}).

Our discussion thus far assumes that the second derivative of the entropy with respect to the energy is not zero.    If we take all our results at face value, then when $d^2s/d\epsilon^2 \rightarrow 0$, the specific heat will become infinite [Eq (\ref{C_and_s})], as will the variance of the energy per particle [Eq (\ref{var_eps})].  This is a critical point.  

There is much more to be said about the analysis of critical points using the entropy vs energy.  But our concern here is how these ideas connect to systems that are not in thermal equilibrium, so that temperature and energy are not relevant concepts.  What we would like to show is that many of the  thermodynamic quantities  nonetheless serve to characterize the behavior of {\em any} probability distribution for a very large number of variables.

Rather than trying to compute the partition function, we can ask, for any distribution, how the normalization condition is satisfied.  We still imagine that there are states $s$, built of of $N$ different variables,  as with the patterns of spiking and silence in a network of neurons.  Each state $s$ has a probability $P_s$, and we must have
\begin{equation}
1 = \sum_s P_s .
\end{equation}
We can now follow the same strategy that we used above for the partition function:  we do the sum first by summing over all the states that have the same value of the (log) probability, and then we sum over this value.  We start by defining 
\begin{equation}
E_s = -\ln \left({{P_s}\over{P_0}}\right),
\end{equation}
where $P_0$ is  the probability of the most likely state, as in Eq (\ref{Edef}).  Then we have
\begin{equation}
\sum_s P_s = \sum_s \int dE \,\delta (E -E_s)  P_s .
\end{equation}
But since $P_s = P_0 e^{-E_s}$, we can rewrite this as
\begin{equation}
\sum_s P_s  = P_0 \int dE \, e^{-E} \sum_s \delta (E -E_s) .
\end{equation}
Integrating by parts, we obtain
\begin{equation}
\sum_s P_s = P_0\int dE \, e^{-E}   {\cal N}(E) ,
\end{equation}
where  ${\cal N}(E)$ is a cumulative density of states, as in Eq (\ref{cumNE}),
\begin{equation}
{\cal N}(E) = \sum_s \Theta  ( E -E_s).
\end{equation}
Again, this is a number of states, so the logarithm of this number is an entropy, exactly as in Eq (\ref{SfromN}).  Thus the statement that the probability distribution is normalized becomes
\begin{equation}
\sum_s P_s = P_0\int dE \,\exp\left[ - E + S(E)\right] .
\label{norm2}
\end{equation}

If we have system in which the state $s$ is built out of $N$ variables, then we expect that, for large $N$, both log probabilities ($E$) and entropies ($S$) are proportional to $N$.  A standard example is in information theory, where $s$ could label a message built out of $N$ symbols, and the proportionality $E \propto N$ is central to proofs of the classic coding theorems \cite{cover+thomas_91}.  In the case of interest to us here, we can look at the states taken on by groups of $N$ neurons, and we can vary $N$ over some range.  The function ${\cal N}(E)$, and hence the entropy $S(E)$, is a property of a single system with a particular value of $N$, and to remind us of this fact we can write $S_N(E)$.  It is an experimental question what happens as $N$ become large.  But, in many of the examples we understand---from statistical physics, from information theory, and indeed from more general examples in probability theory---we find that there is a well defined limiting behavior at large $N$, which means that there is a function
\begin{equation}
s(\epsilon ) = \lim_{N\rightarrow\infty} {1\over N} S_N (E = N\epsilon ).
\end{equation}
If this limit exists, then the normalization condition on the probability distribution in Eq (\ref{norm2}) becomes
\begin{eqnarray}
\sum_s P_s &\rightarrow& N P_0\int d\epsilon \,e^{-Nf(\epsilon )} ,\\
f(\epsilon ) &=& \epsilon - s(\epsilon ) .
\end{eqnarray}
Now we can see the correspondence with the description of an equilibrium thermodynamic system, which leads up to the expressions for the partition function in Eqs (\ref{Zf}) and (\ref{fdef}):

1.  We can assign an ``energy'' to every state of the system, which is just the negative log probability.  It is convenient to normalize so that the most likely state has zero energy.  The ``effective temperature'' of the system is $k_B T = 1$.

2.  We can count the number of states below a given energy, and the log of this number is an entropy.

3.  If there are $N$ elements (e.g., neurons) in our system, it is natural to ask about the entropy per element as a function of the energy per element.  If this function has a smooth limit as $N$ becomes large, $s(\epsilon )$, then we can define a thermodynamics for the system.

4.  When we sum over states, the sum is dominated by states that minimize the free energy, $f(\epsilon ) = \epsilon - s(\epsilon )$, just as in ordinary thermodynamics, provided that the curvature of the free energy at this minimum is nonzero.  

5.  The dominance of states near at the minimum of the free energy enforces the notion of ``typicality'' \cite{cover+thomas_91}, so that at large $N$  most of the states we actually see have essentially the same value of log probability.

6.  If the curvature at the minimum of the free energy vanishes, then the usual ideas of typicality break down, and we will see large fluctuations in the log probability of states, even if we normalize this log probability by $N$.

7.  The large variance in log probability is mathematically equivalent to a diverging specific heat in the thermodynamic case.  This is a signature of a critical point.

Before leaving this discussion, we should note that there are other signatures of criticality, and even different notions of criticality.  In equilibrium systems with interactions that extend only over short distances, correlations typically extend over some longer but finite distance $\xi$; at the critical point this correlation diverges, so that there is no characteristic length scale---all scales between the size of the constituent particles and the size of the system as a whole are relevant \cite{wilson_79}.  Not only does the specific heat diverge at the critical point, but so does the susceptibility to external fields.  All of these diverging quantities have a power--law dependence on the difference between the actual temperature and the critical temperature, and the exponents of these power--laws are quantitatively universal:  many different systems, with different microscopic constituents, exhibit precisely the same exponents, and in a certain precise sense these exponents give a complete description of the system in the neighborhood of the critical point \cite{textbook,parisi_88}.  In the study of complex, non--equilibrium systems, scale invariance and power--law behaviors often are taken as signs of criticality, but seldom is it possible to exhibit these behaviors over the wide range of scales that are the standard in studies of equilibrium critical phenomena, so one must be cautious.   

In almost all equilibrium systems, the approach to criticality also is associated with the emergence of long time scales in the dynamics; as with the divergence of the correlation length $\xi$, the divergence of the correlation time in the dynamics typically means that there is a form of temporal scale invariance at criticality.  Deterministic dynamical systems also exhibit critical phenomena, often called bifurcations, where the system's behavior changes qualitatively in response to an infinitesimal change in parameters \cite{guckenheimer+holmes}.  These phenomena are easiest to understand when the number of degrees of freedom $N$ is small, but then the sharp bifurcations are rounded if there is noise in the system; the example of equilibrium statistical mechanics shows how noisy dynamical systems can recover sharp transitions in the limit of large $N$.  In general it is not clear how dynamical and statistical notions of criticality are related to one another in systems with many degrees of freedom.

\section{Experimental methods}
\label{methods}

Much of the analysis in this paper is based on the same data set as in Ref \cite{tkacik+al_14a}.  For completeness we review our experimental methods here.

Experiments were performed on the larval tiger salamander, {\em Ambystoma tigrinum tigrinum}, in accordance with institutional animal care standards. Retinae were isolated from the eye in darkness, and the retina was pressed against a custom fabricated array of 252 electrodes. The retina was superfused with oxygenated Ringer's medium ($95\% \ {\rm O}_2,\, 5\% \ {\rm CO}_2$) at room temperature. Electrode voltage signals were acquired and digitized at 10 kHz by a 252 channel preamplifier (MultiChannel Systems, Germany). The sorting of these signals into action potentials from individual neurons was done offline using the methods of Ref \cite{marre+al_12}. 

The repeated natural movie was a movie of a fish tank captured at 30 Hz with a standard camera; it lasted 20 s, and was repeated 297 times.  As noted in the text, this experiment allowed us to resolve 160 neurons across the recording array.  The random checkerboard consisted of square pixels, $69\,\mu{\rm m}$ on a side, each chosen independently black or white 30 times per second, creating a 30 second random movie that was repeated 69 times; this experiment yielded 120 stable, resolved cells.   For the spatially uniform flicker, the luminance of the entire screen was chosen randomly from a Gaussian distribution 60 times per second, creating a ten second long random sequence that was repeated 98 times; we separated the signals from 111 neurons.

\section{Density of states and heat capacity in maximum entropy models}
\label{app:DoS1}

We can take our maximum entropy model seriously as a statistical mechanics problem and use Monte Carlo simulation to generates samples of the states $\{\sigma_{\rm i}\}$ drawn from our model distribution.  Heat capacity curves were estimated by running a Metropolis Monte Carlo sampler independently at every $T$.  Since the model assigns an energy $E= {\cal H}(\{\sigma_{\rm i}\})$ to each state, we can compute the mean and variance of $E$ from a single long Monte Carlo run, and thus estimate the heat capacity through the thermodynamic identity in Eq (\ref{CandVarE}).   Samples of the energy were collected at every sweep (roughly $N$ spin flips); $2\times 10^6$ sweeps were performed for every $T$.   

To estimate the function $n(E)$ in the maximum entropy models, including the $\alpha$--ensembles of  Fig \ref{alpha_fig}, we used Wang--Landau sampling \cite{WL}.  In detail, the complete energy range was divided into $2\times 10^4$ equidistant energy bins ($6\times 10^3$ for the $\alpha$--ensembles), the histogram flatness criterion was 0.9, and the final multiplicative update $1+10^{-5}$.  These measurements, as well as the specific heat curves, can both be used to give an estimate of the entropy of the distribution, and these agree to within better than one percent \cite{tkacik+al_14a}, providing a check on our sampling procedures.

For more on these matters, see the methods section ``Computing the entropy ...''  of Ref \cite{tkacik+al_14a}.

\section{More about alternatives}
\label{app:alt}

In this section we expand on alternative interpretations of the data, arguing that the signatures of criticality are unlikely to be explained away as spurious consequences of less interesting models.

\subsection{Impact of limited data}  
\label{limited_data}

\begin{figure}[tb]
\includegraphics[width=\linewidth]{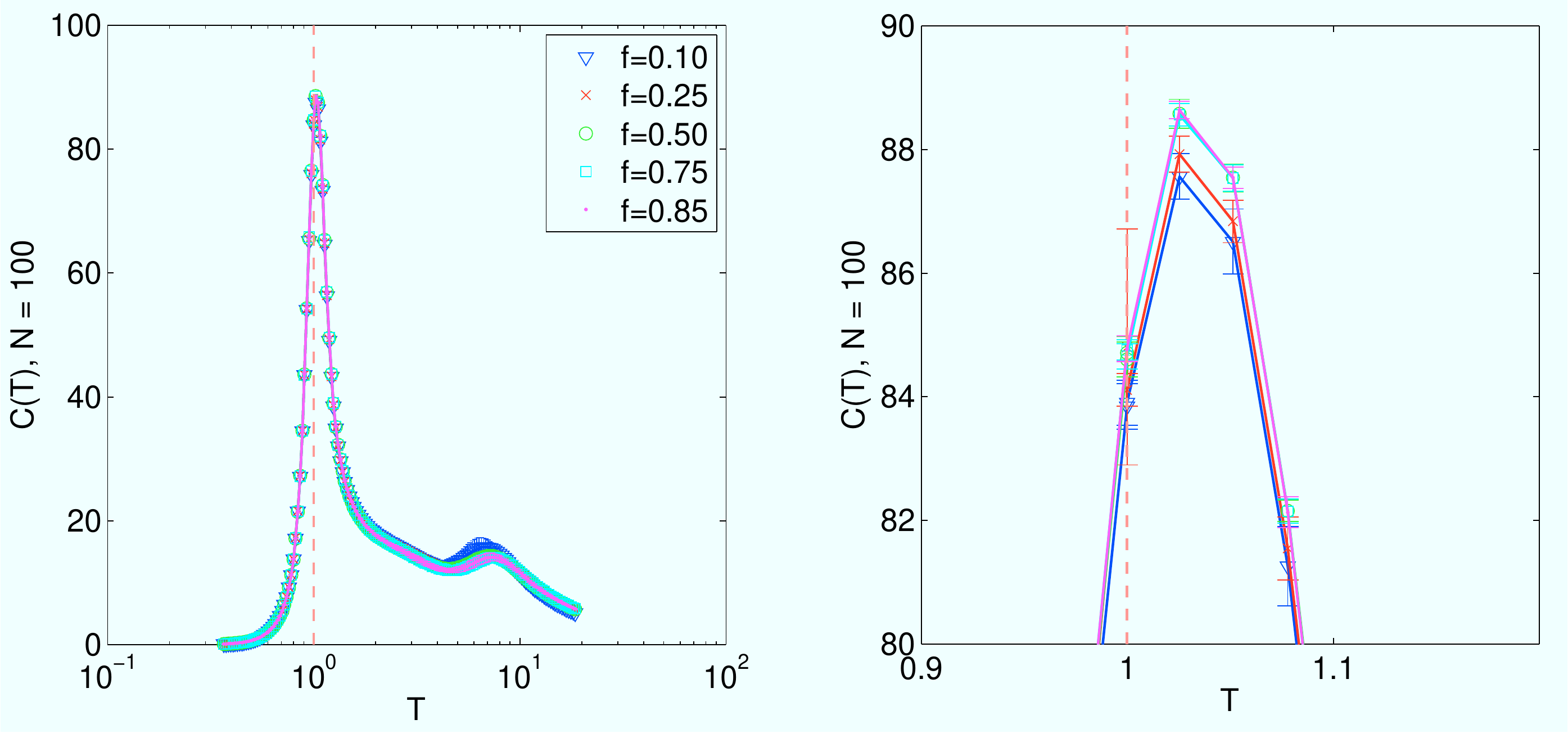}
\caption{Heat capacity in maximum entropy models learned from limited data.  At left, $C(T)$ for models learned from different fractions $f$ of the data.  At right, we zoom in on the peak near $T=1$; error bars show standard deviations across different randomly chosen fractions of the data.  \label{Cv_finitesize}}
\end{figure}

We have tested in detail the reliability with which maximum entropy models can be inferred from the available data.  As explained in Ref \cite{tkacik+al_14a}, we can learn these models from $90\%$ of the data, and then compare the quality of the model against both the training set and the held out test set.  Even with  $N=120$ neurons, the model predicts that the log--likelihood of the test data is the same as that of the training data, within error bars, and these errors are less than $1\%$ (Fig 4 of Ref \cite{tkacik+al_14a}).  Still, one could worry that small errors associated with the finiteness of the data set could have a disproportionate impact on the putative signatures of criticality. To test for this, we have learned models for $N=100$ neurons from fractions of the data ranging down to just $10\%$; results for the heat capacity vs temperature (as in Fig \ref{Cv}) are shown in Fig \ref{Cv_finitesize}.  We see that the sharp peak in $C(T)$ is essentially independent of the sample size across this wide range, and that the variations in $C(T)$ across different small fractions of the data are only a few percent.  Thus, this behavior is {\em not} a result of over fitting, nor is it linked in any way to the size of our data set.

It is important that, in Fig \ref{Cv_finitesize}, we are always looking at the same 100 neurons, else variability across subsets could be confused with sampling errors.  When we change the size of the data we are choosing at random some fraction of the experiment, and for each fraction we examine 10 such random choices.  For each choice we make a completely independent reconstruction of the maximum entropy model, which means that variability includes not just the effects of finite data but also any errors in parameter estimation or in the Monte Carlo estimate of the specific heat.  Evidently {\em all} of these errors are quite small.

\subsection{Are correlations inherited from the visual stimulus?}
\label{cond_ind}

As discussed in the main text, one possible interpretation of our observations is that correlations among neurons simply reflect correlations in the visual stimulus.  In this case, any interesting features in the joint distribution of activity among many neurons would be entirely traceable to the structure of the sensory inputs.  

The idea that correlations among neurons should be decomposed into contributions from their inputs and contributions intrinsic to the circuit is very old \cite{perkel+bullock_68}, dating back to a time when it was hoped that measurement of correlations would allow a direct inference of connectivity in the circuit.  Before discussing the origin of correlations, it is important to emphasize that the distinction between ``stimulus induced'' and ``intrinsic'' correlations is {\em not} a distinction that the brain can make.  Experimentally, we make this distinction by providing exact repetitions of the stimulus, but this never happens in the natural world, and the only knowledge that that brain has of its visual inputs is the set of signals provided by the population of ganglion cells itself, so there is no way to search for correlations with some other reference signal. 

\begin{figure}[t]
\includegraphics[width=\linewidth]{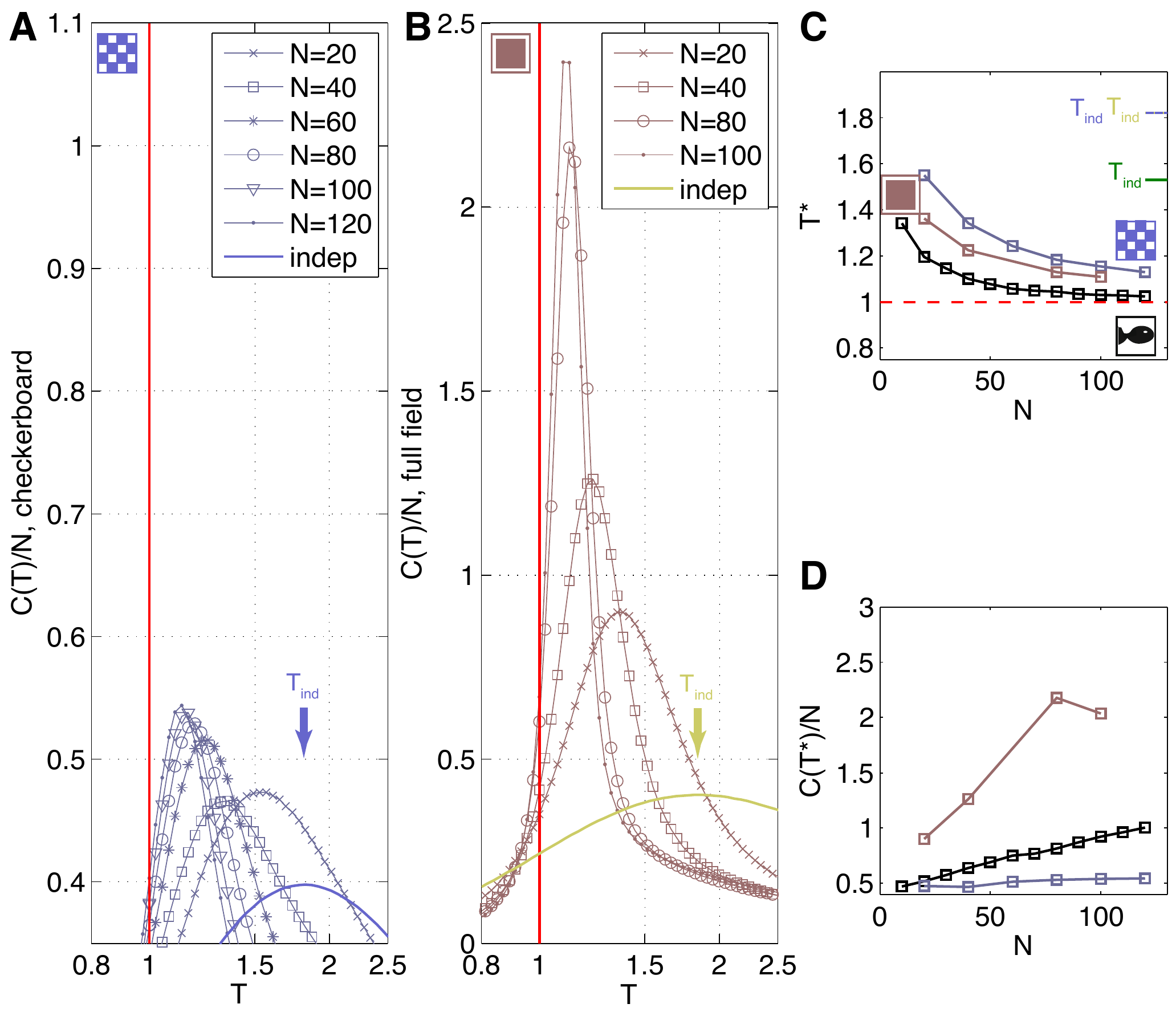}
\caption{{\bf Specific heat in different stimulus ensembles.} {\bf (A)} Random checkerboard stimuli, averaged over five subnetworks at each $N$.  {\bf (B)} Full field flicker stimuli, averaged over 30 subnetworks at each $N$.   In both panels, arrows indicate the peak of heat capacity for a matched network of independent neurons; error bars not shown for clarity. {\bf (C)} The approach of the peak of specific heat, $T^*$, to $T=1$ (model reconstructions), for fish movie (black, from Fig \ref{Cv}), full field flicker (brown), and checkerboard (violet). {\bf (D)} The growth of specific heat with $N$ for the same stimulus ensembles.   \label{compare_ens}}
\end{figure}

While the dissection of the correlations is irrelevant for brain function, it is interesting to ask, mechanistically, how these correlations arise.  If they arise solely from the visual inputs, then changing the statistical structure of these inputs should produce a dramatic effect.  We have replaced the natural movies with random flickering checkerboards (an approximation to spatiotemporal white noise) and spatially uniform but temporally random flicker (Appendix \ref{methods}).  In each case we have constructed maximum entropy models [Eqs (\ref{PH}) and (\ref{HKpair})] and searched for a peak in the specific heat vs temperature, as in Fig \ref{Cv}; results are shown in Fig \ref{compare_ens}.

Although there are quantitative differences among the responses to the different stimulus ensembles, we see that there are signatures of criticality in each case.  As with the natural movies, there is a peak in the specific heat, the height of the peak grows with the number of neurons, and the location of the peak moves toward $T=1$ at larger $N$.  It thus seems unlikely that these signatures of criticality in the specific heat are merely a reflection of input statistics.  Indeed, we should remember that the decomposition of correlations into intrinsic and stimulus--induced is incomplete, because the retina adapts to the distribution of its inputs, on many time scales.  It would appear that some combination of anatomical connectivity and adaptation poises the population of retinal ganglion cells near a peak in the specific heat.  This points toward future experiments that should probe more directly the invariance of thermodynamic behavior across adaptation states \cite{ioffe+al_14}.

\begin{figure}[b]
\includegraphics[width=\linewidth]{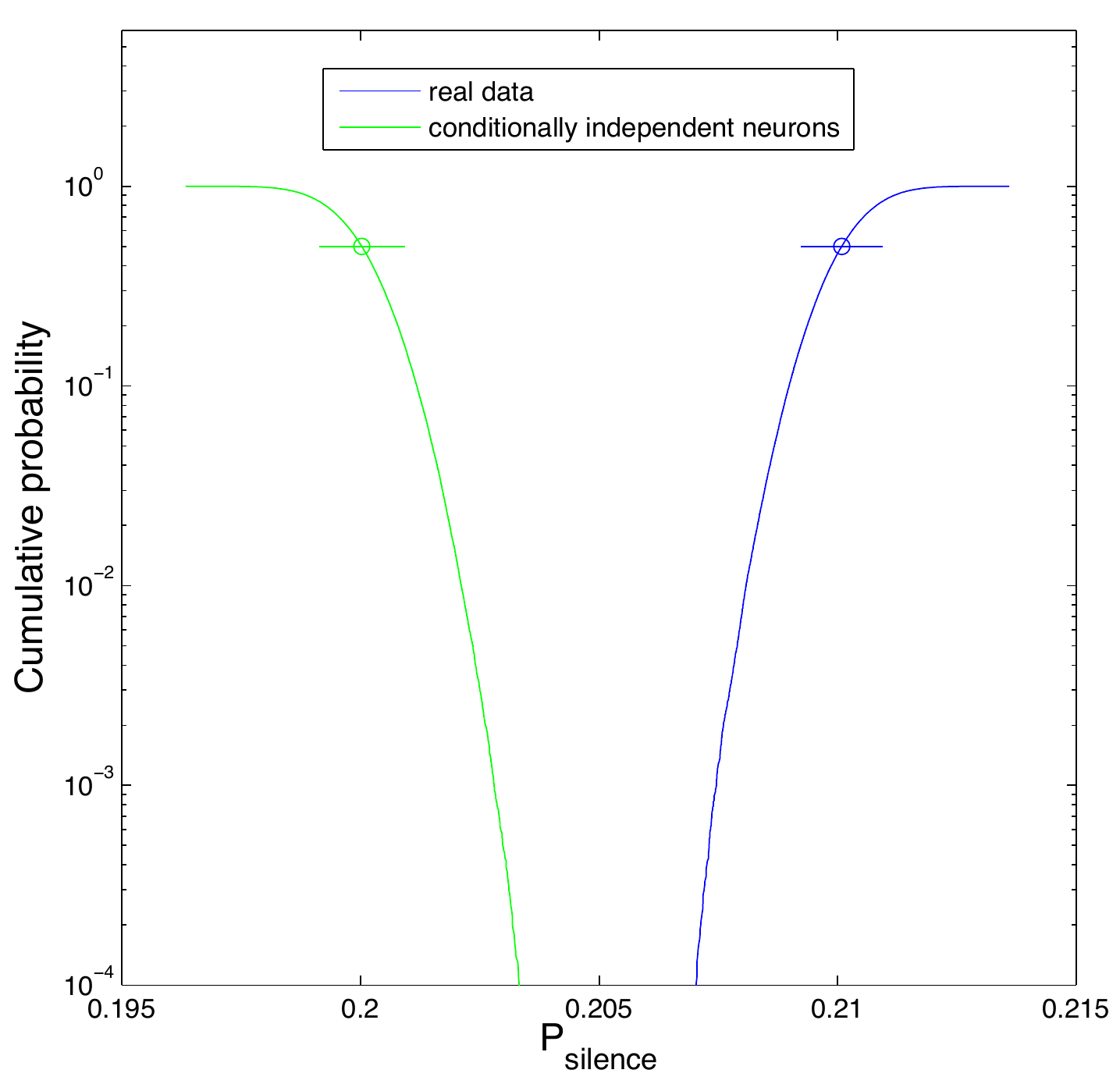}
\caption{The probability of silence.  We compute the probability of the silent state in the full population of $N=160$ neurons, using randomly chosen halves of the data.  Mean and standard deviation as point with error bars, blue for the real data and green for surrogate data in which trials are shuffled to give conditionally independent neurons.  Blue (green) curve shows the probability of finding, in half of the data, a  fraction of silent bins less than (more than) a given value for the real data (conditionally independent neurons).  The two distributions are non--overlapping, to better than $p=10^{-4}$.\label{silence}}
\end{figure}

We can test the picture of stimulus--driven correlations more directly by creating surrogate data in which the neurons respond independently to the visual stimulus.  Since the visual stimuli consist of many repetitions of a short movie clip, we can do this by permuting the labels on the repetitions independently for each neuron.  We can compute many statistical properties of these surrogate data, but our previous analysis \cite{tkacik+al_14a} emphasized that the probability of observing complete silence in the population is a very sensitive test of our models.  As shown in Fig \ref{silence}, the probability of silence in the surrogate data differs from that in the real data by a small amount, but this difference is more than ten times the standard errors in our estimates:  $P_{\rm silence} = 0.2101\pm 0.0009$ for the real data vs $P_{\rm silence} = 0.2000\pm 0.0009$ for the surrogate data.   To make this more precise, we look at the distribution of our estimates over many randomly chosen halves of the data, and find that even if choose $10^5$ times we never find  the probability of silence in the surrogate data to be as large as in the real data. Thus we can reject the conditionally independent model with high confidence.  

\subsection{More general hidden variable models}
\label{SNM}

The idea that correlations among neurons might be inherited from the visual stimulus is one possibility among many.  More generally we might ask if the pattern of correlations could be understood as the independent response of neurons to {\em some} signal that is effectively external to the network, or at least hidden from an observer that sees only the patterns of spikes and silence.  To assess this possibility it is useful to step back and think about well known models in statistical mechanics.

Consider the mean--field Ising ferromagnet, in which spins $\sigma_{\rm i} = \pm 1$ experience an effective magnetic field that is proportional to the average over all the other spins in the system, so that
\begin{equation}
E(\{\sigma_{\rm i}\}) = -{J\over{2N}}\sum_{{\rm i}\neq {\rm j}} \sigma_{\rm i}\sigma_{\rm j} .
\label{MF1}
\end{equation}
Note that the sum is over all pairs, and the factor of $N$ insures that the energy of the system is proportional to $N$.   The sum over all distinct pairs is missing the term ${\rm i} = {\rm j}$, but since $\sigma_{\rm i}^2 = 1$ we have
\begin{eqnarray}
E(\{\sigma_{\rm i}\}) &=& -{J\over{2N}}\sum_{{\rm i},  {\rm j}} \sigma_{\rm i}\sigma_{\rm j}  + {J\over 2}\\
&=& -{J\over{2N}} \left( \sum_{\rm i} \sigma_{\rm i}\right)^2 + {J\over 2} .
\end{eqnarray}
Thus when we compute the partition function, we have (choosing units where $k_B T = 1$)
\begin{eqnarray}
Z &\equiv& \sum_{\{\sigma_{\rm i}\}} e^{-E(\{\sigma_{\rm i}\}) /k_B T}\\
&=& \sum_{\{\sigma_{\rm i}\}} \exp\left[ {J\over{2N }} \left( \sum_{\rm i} \sigma_{\rm i}\right)^2 - {J\over{2 }}\right] \label{Z4}
\end{eqnarray}
But we can always write
\begin{equation}
\exp\left[ {A\over 2}x^2 \right]  = \int {{dh}\over{\sqrt{2\pi A}}}  \,\exp\left[ -{1\over {2A}} h^2 + h x\right] .
\end{equation}
Applying this identity to Eq (\ref{Z4}), we have
\begin{equation}
Z = e^{-J/2 } \int {{dh}\over{\sqrt{2\pi N/J}}} \sum_{\{\sigma_{\rm i}\}} \exp\left[ -  {{N}\over{2J}} h^2 + h \sum_{\rm i} \sigma_{\rm i}\right] .
\end{equation}
We can think of this, more suggestively, as
\begin{eqnarray}
Z &=& e^{-J/2} \int {dh} \,P(h)  \sum_{\{\sigma_{\rm i}\}} \exp\left[  h \sum_{\rm i} \sigma_{\rm i}\right] \\
&=& e^{-J/2} \int {dh} \,P(h)  \sum_{\{\sigma_{\rm i}\}} e^{-E(\{\sigma_{\rm i}\}; h)},
\label{Zh1}
\end{eqnarray}
where $P(h)$ is a Gaussian probability distribution for $h$, with zero mean and variance $J/N$.  But now the sum over states involves a model in which each spin $\sigma_{\rm i}$ experiences  a magnetic field $h$, so that the total energy
\begin{equation}
E(\{\sigma_{\rm i}\}; h) = -h\sum_{\rm i}  \sigma_{\rm i} .
\label{Eh1}
\end{equation}
Thus a model in which all spins interact with one another, equally, is {\em mathematically identical} to a model in which each spin responds independently to a magnetic field chosen at random from a Gaussian distribution.

We can find essentially the same equivalence in a broader class of models, which includes the Hopfield model of associative memories \cite{hopfield_82}.  Consider a collection of spins that interact through some matrix $J_{\rm ij}$, so that the energy
\begin{equation}
E(\{\sigma_{\rm i}\}) = {1\over 2} \sum_{{\rm i},{\rm j}} J_{\rm ij}\sigma_{\rm i}\sigma_{\rm j} .
\label{Jij}
\end{equation}
The Hopfield model corresponds to the choice
\begin{equation}
J_{\rm ij} = - {J\over N}\sum_{\mu =1}^K \xi_{\rm i}^\mu \xi_{\rm j}^\mu ,
\end{equation}
where there are $K$ ``stored memories'' 
\begin{equation}
{\mathbf\xi}^\mu \equiv \{\xi_1^\mu ,\, \xi_2^\mu,\,\cdots ,\, \xi_N^\mu\}.
\end{equation}
In this case, the same arguments that lead to Eqs (\ref{Zh1}) and (\ref{Eh1}) now give
\begin{equation}
Z \propto {\bigg\langle} \sum_{\{\sigma_{\rm i}\}} e^{-E(\{\sigma_{\rm i}\}; \{h_{\rm i}\})} {\bigg\rangle} ,
\label{Zh2}
\end{equation}
where the energy corresponds to each spin responding independently to a magnetic field,
\begin{equation}
E(\{\sigma_{\rm i}\}; \{h_{\rm i}\}) = -\sum_{\rm i} h_{\rm i}\sigma_{\rm i} ,
\label{Eh2}
\end{equation}
and the magnetic fields are
\begin{equation}
h_{\rm i} = \sum_{\mu =1}^K \phi_\mu \xi_{\rm i}^\mu ,
\label{Kfields}
\end{equation}
where each $\phi_\mu$ is a Gaussian random variable with zero mean and variance $\langle\phi_\mu^2\rangle = J/N$; the average in Eq (\ref{Zh2}) is an average over these fluctuating fields.

In fact this construction is yet more general.  If the energy has the form of Eq (\ref{Jij}), and the matrix $-J_{\rm ij}$ has all positive eigenvalues, then we can rewrite the partition function as an average over a model of independent spins in magnetic fields, as in Eq (\ref{Eh2}), where the fields are Gaussian random variable with zero mean and a covariance matrix $\langle h_{\rm i} h_{\rm j}\rangle = (-J^{-1})_{\rm ij}$.

The conclusion from these arguments, which are well known, is that a large class of models for interacting spins are mathematically equivalent to models of spins that respond independently to fluctuating magnetic fields.  As applied to models for the activity of neurons, this means that large classes of models for correlated activity are {\em identical} to models of conditionally independent neurons responding to fluctuating inputs.  

How do these identities relate to the signatures of criticality?   The mean--field model defined by Eq (\ref{MF1}) has a critical point at $J=1$, which means that in Eq (\ref{Zh1}) the variance of the field $h$ must be tuned to exactly $1/N$ to achieve a critical state.  In this sense, saying that the system is equivalent to independent spins experiencing a random field doesn't ``explain'' anything, it just moves the problem from the distribution of states being poised at a special point to the distribution of fields being poised at some special point.

Recently, Schwab et al \cite{schwab+al_13} have suggested a different view.  If we think of the effective magnetic field as arising from a truly external source, then instead of being generated internally, then it is more natural to write
\begin{eqnarray}
P(\{\sigma_{\rm i}\}) &=& \int dh\, P(h) P(\{\sigma_{\rm i}\} | h)\\
& =& \int dh\, P(h) e^{F(h)} e^{-E(\{\sigma_{\rm i}\}; h)} ,
\label{D18}
\end{eqnarray}
where the free energy $F(h) = -N\ln\left[ 2\cosh(h)\right]$, and the energy $E(\{\sigma_{\rm i}\}; h)$ as in Eq (\ref{Eh1}).  Note the difference between this equation and Eq (\ref{Zh1}):  because we take the field as being truly external, there is an extra factor of the (exponential of) the free energy in the integral, and this is a huge effect, because the free energy is proportional to $N$.  When the fields are generated internally, as a consequence of the interactions, then in the simple case of Eq (\ref{Zh1}) we find that the variance of the field is fixed to be $J/N$, and hence the fields are small.  In contrast, in Eq (\ref{D18}), we are free to imagine any distribution that we might like.

In particular, if the distribution of fields $P(h)$ is sufficiently broad, then the structure of this distribution plays relatively little role in the overall distribution of states.  Mathematically, if $P(h)$ is broad then the integral over $h$ can be approximated for large $N$ as
\begin{eqnarray}
P(\{\sigma_{\rm i}\}) &\propto& {{P(h=h_*(m))} \over {\sqrt{N (1-m^2)}}} e^{-Ns(m)}
\label{D19}\\
s(m) & =& - {{1+m}\over 2} \ln\left({{1+m}\over 2}\right) \nonumber\\
&&\,\,\,\,\,\,\,\,\,\,\,\,\,\,\,- {{1-m}\over 2} \ln\left({{1-m}\over 2}\right),
\end{eqnarray}
where the magnetization $m= (1/N) \sum_{\rm i}\sigma_{\rm i}$ and $h_*(m)$ is defined by
\begin{equation}
m = \tanh [h_*(m)] .
\end{equation}
At large $N$, Eq (\ref{D19}) tells us that the energy per neuron associated with any state $\{\sigma_{\rm i}\}$ is just $\epsilon = s(m)$, depending only on the magnetization $m$.  But then we can count the number of states with energy near $\epsilon$ by counting the number of states with magnetization $m$, and this shows that $s(m)$ is also the entropy per neuron, so that $s(\epsilon ) = \epsilon$, a signature of criticality.   Could this, or some generalization, provide an explanation for what we see in the data?  

Passing from Eq (\ref{D18}) to Eq (\ref{D19}) hinges on the fact that we are integrating a function of one variable that has a factor of $N$ in the exponential.  We could generalize to have $K$ different fluctuating fields, as in Eq (\ref{Kfields}), but for the large $N$ approximation to be valid we must have $N\gg K$.  In particular, in the general case where we try to simulate an arbitrary matrix of pairwise interactions, we need $N$ distinct fields, and now large $N$ no longer helps us do the integral.  The argument of Ref \cite{schwab+al_13} thus requires that the number of effective fields be small compared to the number of neurons.  If we imagine that the unobserved fields are generated by inputs to the retinal ganglion cells from other cells in the retina, then as we change the complexity of the visual stimulus we expect that the effective dimensionality of these inputs also will change.  Nonetheless, as described in Appendix \ref{cond_ind}, we see signatures of criticality for stimulus ensembles including full field flicker, natural movies, and spatiotemporal white noise.

In models where the neurons do not really interact, but instead respond to a fluctuating field with a large dynamic range, the correlations between neurons have a strength that is set by this dynamic range.  But the argument of Ref \cite{schwab+al_13} is that criticality emerges once the dynamic range is large enough, independent of its precise value.  The appearance of a critical relation between entropy and energy in this class of models thus is decoupled from the strength of correlations in the network.  But Fig \ref{alpha_fig} shows that, in an accurate model for the distribution of activity patterns, if we change one parameter to modulate the strength of correlations while holding the mean spike probabilities fixed, the entropy/energy relation moves away from its critical form whether we increase or decrease the strength of correlations.

\end{document}